\documentclass[aps,prl,reprint,superscriptaddress,floatfix]{revtex4-1}
\usepackage{graphicx} 
\usepackage{color,dcolumn,bm,amssymb,amsmath} 
\usepackage{soul,xcolor} 
\usepackage{siunitx} 
\usepackage{epstopdf,gensymb} 
\usepackage{hyperref} 
\DeclareSIUnit\angstrom{\text{\AA}} 
\usepackage{braket}

\def\nactive{N_{\text{active}}}
\def\nflav{N_{\text{flavors}}}
\def\ndeg{N_{\text{deg.}}}

\def\dens{{n}}

\newcommand{\ketbra}[1]{\ket{#1}\bra{#1} }
\newcommand{\cth}{C_{3z}}

\newcommand{\angstrom}{\dot{A}}

\newcommand{\proj}{\mathcal P_{\text{active}} }

\DeclareUnicodeCharacter{03BD}{{$\nu$}}
\begin{document}
\title{Flat band surface state superconductivity in thick rhombohedral graphene}
\author{Yi Guo}
\affiliation{Department of Physics, University of California at Santa Barbara, Santa Barbara CA 93106, USA}
\author{Owen I. Sheekey}
\affiliation{Department of Physics, University of California at Santa Barbara, Santa Barbara CA 93106, USA}
\author{Trevor Arp}
\affiliation{Department of Physics, University of California at Santa Barbara, Santa Barbara CA 93106, USA}
\author{Kry\v{s}tof Kol\'a\v{r}}
\affiliation{Department of Applied Physics, Aalto University School of Science, FI-00076 Aalto, Finland}
\author{Thibault Charpentier}
\affiliation{Department of Physics, University of California at Santa Barbara, Santa Barbara CA 93106, USA}
\author{Ludwig Holleis}
\affiliation{Department of Physics, University of California at Santa Barbara, Santa Barbara CA 93106, USA}
\author{Ben Foutty}
\affiliation{Department of Physics, University of California at Santa Barbara, Santa Barbara CA 93106, USA}
\author{Aidan Keough}
\affiliation{Department of Physics, University of California at Santa Barbara, Santa Barbara CA 93106, USA}
\author{Maya Kang-Chou}
\affiliation{Department of Physics, University of California at Santa Barbara, Santa Barbara CA 93106, USA}
\author{Martin E. Huber}
\affiliation{Departments of Physics and Electrical Engineering, University of Colorado Denver; Denver, Colorado
80217, USA}
\author{Takashi Taniguchi}
\affiliation{International Center for Materials Nanoarchitectonics,
National Institute for Materials Science,  1-1 Namiki, Tsukuba 305-0044, Japan}
\author{Kenji Watanabe}
\affiliation{Research Center for Functional Materials,
National Institute for Materials Science, 1-1 Namiki, Tsukuba 305-0044, Japan}
\author{Cyprian Lewandowski}
\affiliation{National High Magnetic Field Laboratory, Tallahassee, FL 32310, USA}
\affiliation{Department of Physics, Florida State University, Tallahassee, FL 32306, USA}
\author{Andrea F. Young}
\email{andrea@physics.ucsb.edu}
\affiliation{Department of Physics, University of California at Santa Barbara, Santa Barbara CA 93106, USA}
\date{\today}

\begin{abstract}
Rhombohedral multilayer graphene has recently emerged as a rich platform for studying correlation driven magnetic, topological and superconducting states\cite{chen_evidence_2019,
chen_tunable_2020,shi_electronic_2020,kerelsky_moireless_2021,zhou_half-_2021, zhou_superconductivity_2021,zhou_isospin_2022,zhang_enhanced_2023,
han_orbital_2023, liu_spontaneous_2023,zhou_layer-polarized_2024,holleis_nematicity_2025,zhang_twist-programmable_2025,han_signatures_2025,auerbach_isospin_2025,holleis_fluctuating_2025, zhang_layer-dependent_2025, li_transdimensional_2025, deng_superconductivity_2025, kumar_superconductivity_2025, morissette_superconductivity_2025,yang_magnetic_2025}.
While most experimental efforts have focused on devices with N$\leq 9$ layers, the electronic structure of thick rhombohedral graphene features flat-band surface states even in the infinite layer limit\cite{otani_intrinsic_2010,heikkila_dimensional_2011,xiao_density_2011}. 
Here, we use layer resolved capacitance measurements\cite{young_capacitance_2011,young_electronic_2012,hunt_direct_2017,zibrov_tunable_2017} to directly detect these surface states for $N\approx 13$ layer rhombohedral graphene devices. 
Using electronic transport and local magnetometry, we find that the surface states host a variety of ferromagnetic phases, including both valley imbalanced quarter metals and broad regimes of density in which the system spontaneously spin polarizes. 
We observe several superconducting states localized to a single surface state.  
These superconductors appear on the unpolarized side of the density-tuned spin transitions, and show strong violations of the Pauli limit\cite{clogston_upper_1962,chandrasekhar_note_1962} consistent with a dominant attractive interaction in the spin-triplet, valley-singlet pairing channel. 
In contrast to previous studies of rhombohedral multilayers, however, we find that superconductivity can persist to zero displacement field where the system is inversion symmetric.  
Energetic considerations suggest that superconductivity in this regime is described by the existence of two independent surface superconductors coupled via tunneling through the insulating single crystal graphite bulk. 
\end{abstract}
\maketitle

\begin{figure*}[ht!]
\centering
\includegraphics[width=\textwidth]{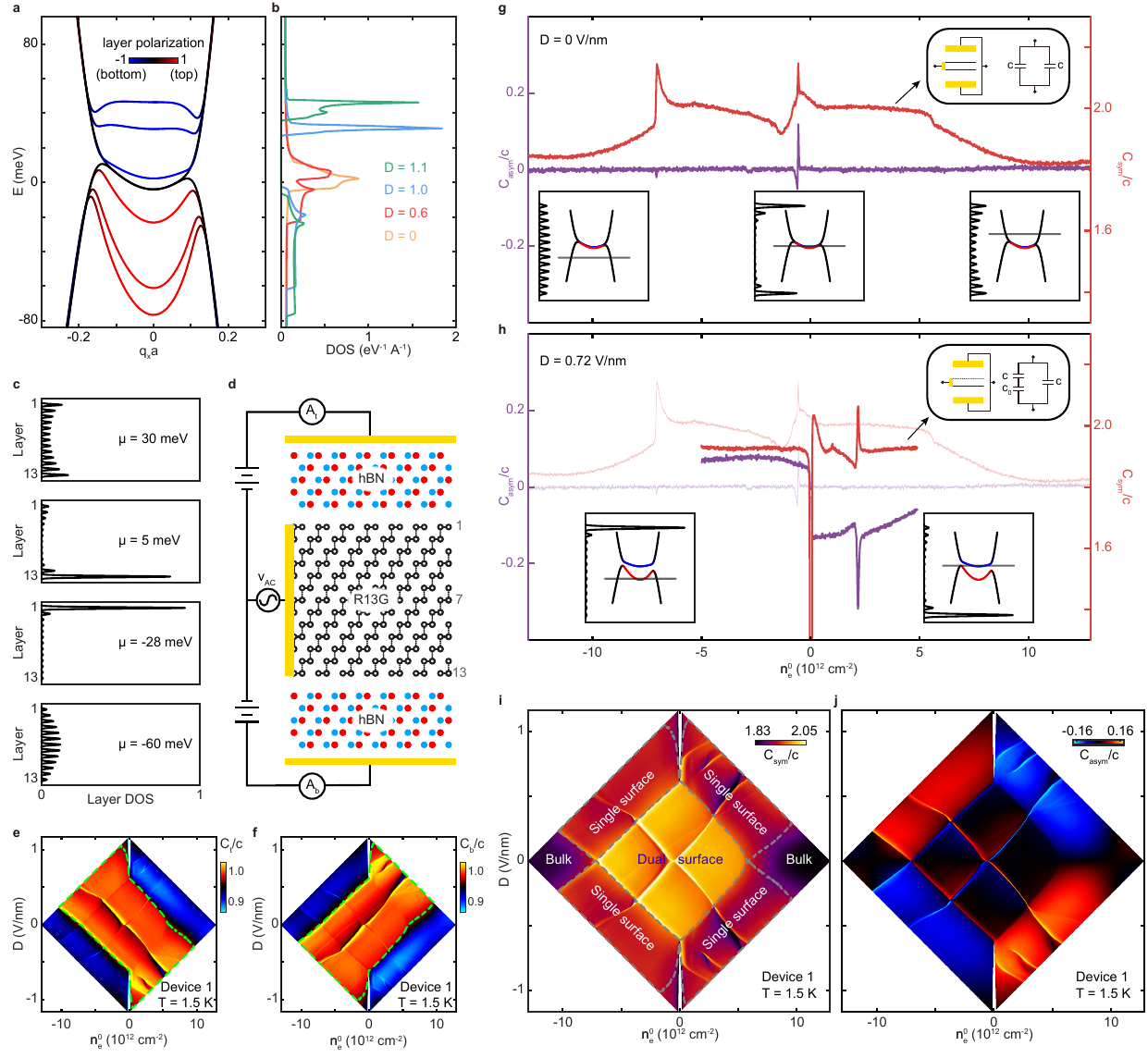}
\caption{
\textbf{Surface states in multilayer rhombohedral graphite.}
\textbf{(a)} 
Band structure and 
\textbf{(b)} density of states for rhombohedral 13-layer graphene calculated within the self-consistent Hartree approximation at charge neutrality for different values of $D$. 
\textbf{(c)} Layer resolved density of states for $D = 0.73 \,\text{V/nm}$ and $\mu = 30$, $5$, $-28$, and $ -60 \,\text{meV}$.
\textbf{(d)} Measurement schematic for layer resolved capacitance.  An AC voltage is applied to the R13G through an electrical contact and the current on the top ($A_t$) and bottom ($A_b$) are measured with  $C_{t(b)}\equiv A_{t(b)}/V_{AC}$. 
\textbf{(e)} Measured $C_t$ and 
\textbf{(f)} $C_b$ at T=1.5K in Device 1, normalized by the geometric capacitance $c$ of the two gates. Between the dashed green lines, $C_{t(b)}\gtrapprox c$, indicating the presence of a high compressibility surface state on the top (bottom) layer. 
\textbf{(g)} $C_{sym}$ and $C_{asym}$ at $D=0$. Lower insets show Hartree band structure and calculated layer density of states for $\mu=0$, $\pm 30 \,\text{meV}$ for $D=0.05$~V/nm.  The upper inset shows an equivalent circuit schematic in the dual-surface regime.
\textbf{(h)}
$C_{sym}$ and $C_{asym}$ at $D=0.72 \,\text{V/nm}$. 
\textbf{(i)} $C_{sym}$ as a function of $n_e^0$ and $D$.
Regions of single-surface, dual-surface, and bulk layer polarization are indicated.  
\textbf{(j)} $C_{asym}$ as a function of $n_e^0$ and $D$. In a simplified four-plate capacitor model, $C_{asym}\propto dp/dn_e^0$; red and blue regions thus denote where states at the Fermi level are top or bottom layer polarized.
\label{fig:fig1}}
\end{figure*}


In an idealized picture, the  unit cell of multilayer rhombohedral (i.e., ABC-stacked) graphite can be described as a dimerized chain in which pairs of covalently bonded carbon atoms in each layer are coupled via an interlayer hopping that is an order of magnitude weaker\cite{mcclure_electron_1969,min_electronic_2008,otani_intrinsic_2010,slizovskiy_films_2019,heikkila_dimensional_2011}. 
At the Brillouin zone corners, the resulting electronic structure is analogous to a Su-Schrieffer-Heeger model\cite{su_solitons_1979} with an even number of lattice sites and dominant odd-numbered bonds, guaranteeing the existence of a state localized at the chain boundaries.  
In the graphene context, this guarantees that the energy bands at the zone corners are polarized on the outermost layers--i.e., they describe surface states even in a crystal of macroscopic thickness. 
However, the perfect surface polarization is spoiled by momentum dependent hopping between adjacent layers as well as hoppings between nonadjacent layers\cite{koshino_interlayer_2010,koshino_trigonal_2009, slizovskiy_nonlinear_2012}. 
This raises the question of the range over which surface states are relevant experimentally, and what their properties might be.

Theoretical work has emphasized the possibility of magnetism and high temperature superconductivity in these systems\cite{kopnin_high-temperature_2011,kopnin_surface_2014,pamuk_magnetic_2017}.  These proposals are particularly salient in light of a body of recent experimental work which has revealed a startlingly similar phase diagram of competing magnetic and superconducting phases in rhombohedral bilayers\cite{zhou_isospin_2022,zhang_enhanced_2023,holleis_nematicity_2025,zhang_twist-programmable_2025}, trilayer\cite{zhou_half-_2021,zhou_superconductivity_2021}, and  4\cite{liu_spontaneous_2023,han_signatures_2025,auerbach_isospin_2025}, 5\cite{han_orbital_2023,han_signatures_2025}, 6\cite{morissette_superconductivity_2025, deng_superconductivity_2025}, 7\cite{zhou_layer-polarized_2024} and 8\cite{kumar_superconductivity_2025} and 9\cite{li_transdimensional_2025} layer rhombohedral graphene devices. 
One early experimental work studied rhombohedral graphite layers as thick as 10 or 20 layers\cite{shi_electronic_2020}, finding experimental evidence for magnetism in 10 layer samples, while recent work has used angle resolved photoemission spectroscopy to visualize flat band surface states in a 50-70 layer sample\cite{zhang_correlated_2024}. 
However, the low-temperature phase diagram of thicker rhombohedral layers has not been explored. 

Here, we study several rhombohedral graphite samples of 13-14 layer thickness, a regime where high energy subbands arising from finite thickness remain irrelevant at low energies but the low energy band structure nevertheless becomes nearly insensitive to layer number\cite{slizovskiy_films_2019}.  
Using layer resolved  capacitance measurements, we quantify the extent of surface polarization, revealing that at low densities nearly all of the electronic states near the Fermi level are localized on either the top or bottom surface. 
These surface states show signatures of superconductivity in the immediate vicinity of thermodynamic phase transitions which local magnetometry measurements reveal to coincide with the onset of spin polarization on a single surface. 
Intriguingly, we find that dual surface superconductivity appears even in absence of an applied electric displacement field, paving the way for the study of Josephson junctions in which the tunnel barrier is composed of the pristine---and insulating---bulk of single crystal graphite.  

\section{Direct detection of flat band surface states}

Figure \ref{fig:fig1}a shows the electronic structure of 13-layer rhombohedral graphene (`R13G') in the vicinity of the Brillouin zone corner located at momentum $\vec K$, calculated within the Hartree approximation for a charge neutral system. 
Bands are color coded by the out-of-plane electric dipole moment per carrier,
$p/n_e=\sum_{i=1}^{N}|\psi_i|^2\frac{2i-N-1}{N-1}$, where $N$ is the number of layers. 
$p/n_e=\pm1$ corresponds to complete polarization on the top- or bottom layer. 
In the absence of an applied displacement field $D \equiv (c_t v_t-c_b v_b)/2\epsilon_{0}=0$ (here $c_t$ and $c_b$ are the geometric capacitance per unit area of the top- and bottom gates, respectively), R13G is a semimetal, with the valence and conduction bands nearly degenerate over a wide range of $\vec q$ near $\vec K$. 
However, at finite $D$, states near the $K$-points become nearly perfectly polarized on the opposite surfaces.
As $D$ increases, this near perfect layer polarization leads to a rapid splitting of the valence and conduction bands, and the eventual opening of a band gap at charge neutrality. 

Our band structure calculations account for the electrostatic energy arising from charging the interlayer capacitors and minimize the resulting total energy with respect to 13 different layer-specific on-site energies. 
In turn, these on-site energies allow different layers to (non-linearly) screen the applied $D$\cite{koshino_interlayer_2010}.  
Our calculations result in on-site potentials that are nearly constant for all but the two surface layers, leading to significantly flatter band dispersions than expected if the potential is screened linearly by the layers. 
This is most evident at intermediate values of $D$ in Fig. 1b, where the conduction band features a flat dispersion and corresponding large density of states. Notably, the peak density of states is comparable to the largest values expected in thinner rhombohedral multilayers\cite{pamuk_magnetic_2017}. 

Fig. 1c shows the layer-resolved DOS (defined as the fraction of the total density of states localized to each layer) at four different values of the chemical potential for $D = 0.73 \,\text{V/nm}$. The four panels correspond to distinct regimes of layer polarization. At strong electron- or hole-doping, the Fermi level is deep in the layer-unpolarized dispersive bands, and electronic density of states is spread almost uniformly across all layers.  In contrast, for light electron- or hole-doping the density of states is concentrated on the outermost surfaces.
To experimentally detect the existence of the surface states, we measure the capacitance $C_{t(b)}$ between the R13G and the top and bottom graphite gate, respectively.  The experimental setup is depicted schematically in Fig. \ref{fig:fig1}d (with details presented in Fig. \ref{fig:SI_CapCircuit}.  In essence, a small finite-frequency excitation voltage $V_{AC}$ is applied to the R13G and the response currents on the two gates ($A_t$ and $A_b$) measured independently\cite{zibrov_tunable_2017}. We note that the sample was fabricated using matched hexagonal boron nitride top and bottom gate dielectrics cut from the same crystal, ensuring that the geometric capacitances of top and bottom gate are equal, $c_t=c_b\equiv c$. 
Figs. \ref{fig:fig1}e and  \ref{fig:fig1}f show $C_t/c$ and $C_b/c$ plotted as a function of $D$ and the nominal electron density, $n_e^0=c(v_t+v_b)$. Consequently, $C_{t(b)}/c \approx 1$---rendered in red on the color scale in Figs. 1e-f---corresponds to a high density of states on the surface nearest a given gate. Four broad regimes are visible. At high $|n_e^0|$ and $D\approx0$, both $C_t$ and $C_b$ are well below this ideal geometric value, indicating that the density of states of the rhombohedral layer is localized away from the surfaces. For small to moderate $|n_e^0|$ and $D$, both $C_t$ and $C_b$ are close to $c$, indicating high density of states on both surfaces.  Finally for large $n_e\cdot D$ product, one capacitance is close to $c$ while one is depressed, with the choice depending on the sign of $n_e\cdot D$--indicating near-complete polarization of the density of states on a single surface. 

\begin{figure*}[t]
\centering
 \includegraphics[width=\textwidth]{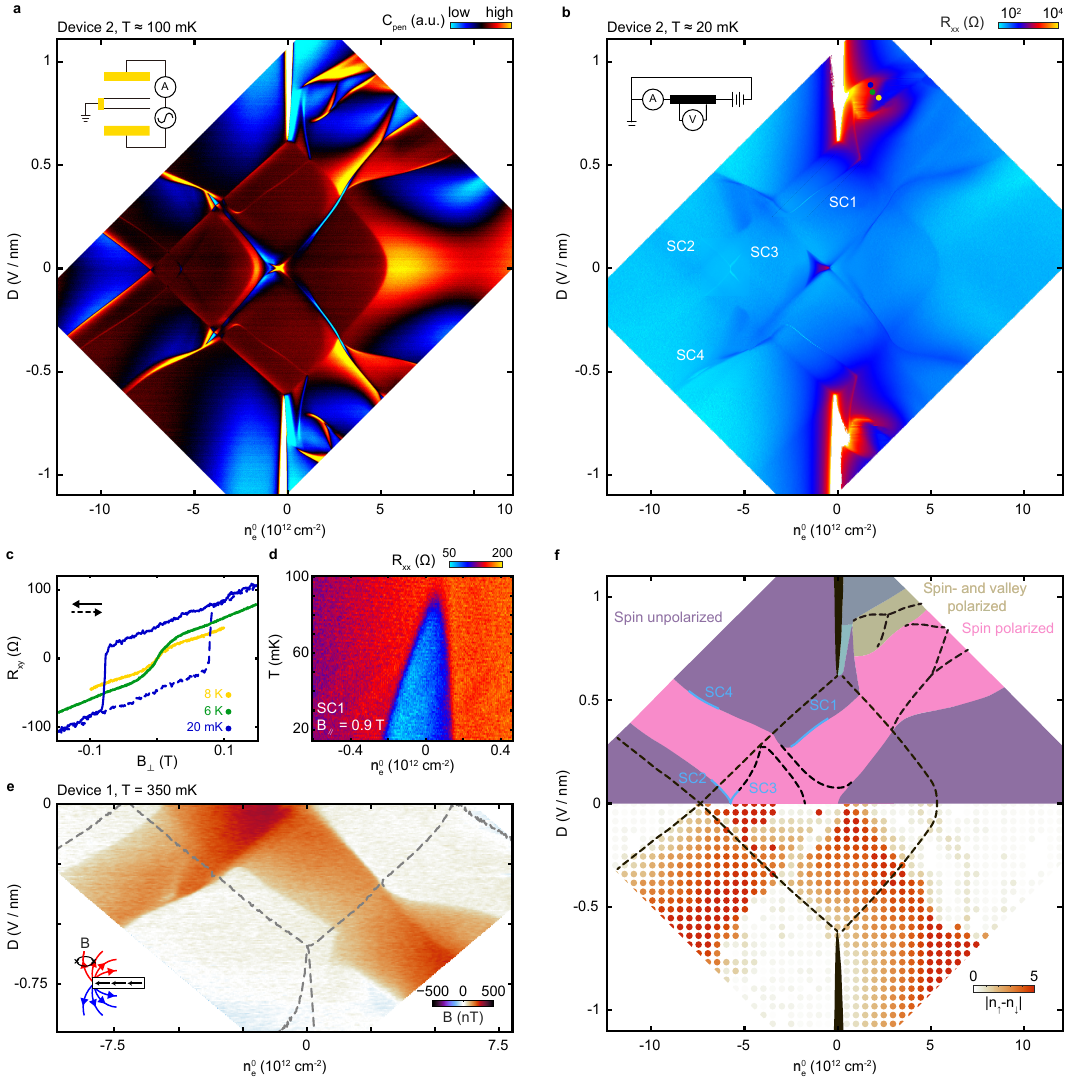}
\caption{\textbf{Spin and valley ferromagnetism and surface state superconductivity.}
\textbf{(a)} Penetration field capacitance of Device 2 measured at T=100mK.  
\textbf{(b)} Longitudinal resistance $R_{xx}$ in Device 2 measured at $T=20\,\text{mK}$.  Several superconducting states are indicated. 
\textbf{(c)} Hall measurements at $20\,\text{mK}$ (blue), $6\,\text{K}$ (green) and $8\,\text{K}$ (yellow). The three curves are measured at $(n_e^0,D,T)=(1.77 \times 10^{12}\,\text{cm}^{-2}, 0.89\,\text{V/nm}, 20\,\text{mK})$, $(n_e^0,D,T)=(1.87 \times 10^{12}\,\text{cm}^{-2}, 0.85\,\text{V/nm}, 6\,\text{K})$ and $(n_e^0,D,T)=(2.21 \times 10^{12}\,\text{cm}^{-2}, 0.83\,\text{V/nm}, 8\,\text{K})$, indicated by the colored dots in panel b.
\textbf{(d)} Temperature dependent $R_{xx}$ in SC1, measured at $D = 0.47\,\text{V/nm}$ and $B_\parallel=0.9T$.  
\textbf{(e)} Fringe magnetic field arising from in-plane magnetization measured by SQUID-on-tip microscopy at T=350mK.  As shown in the inset, the out-of-plane fringe field $B$ is measured over a sample edge, once with  applied external field   $(B_\parallel,B_\perp)=(20 \,\text{mT},22\,\text{mT})$ and once with $(B_\parallel,B_\perp)=(-20 \,\text{mT},20\,\text{mT})$; the plotted data is the difference between the two measurements. 
Overlays (gray dashed lines) indicate single-surface, dual-surface, and bulk regions extracted from Fig. \ref{fig:fig1}e and f. 
\textbf{(f)} Top half: schematic phase diagram of symmetry breaking and superconductivity.  Spin unpolarized phases are indicated in purple; spin polarized phases in pink, and valley polarized phases in tan. 
Superconductors are rendered in cyan and labeled.  The cyan region at low $n_e^0$ and intermediate $D$ has high resistivity and large negative compressibility, and may be a Wigner crystal like state. Bottom half shows the spin imbalance density calculated within a simplified two-flavor Hartree-Fock model described in the methods. 
}
\label{fig:fig2}
\end{figure*}

To understand the emergence of surface states quantitatively, we analyze the symmetric and antisymmetric gate capacitance, defined as $C_{sym (asym)} \equiv C_t \pm C_b$. 
These quantities are most easily understood within a four-plate capacitor model\cite{young_capacitance_2011,young_electronic_2012,zibrov_tunable_2017,de_la_barrera_cascade_2022} where the electrons are assumed to reside only on the outer layers, separated by a dielectric graphite bulk with capacitance $c_0$. 
Within this picture, $C_{sym}/c = 2 \frac{dn}{dn_e^0}$ measures the total capacitance from the gates to the electron system and is is sensitive to both geometric and quantum capacitance effects. $C_{asym}/c= \frac{c}{c_0} \frac{dp}{dn_e^0}$ measures the change in layer polarization with $n_e^0$, corresponding approximately to the layer polarization at the Fermi level. 

Fig. \ref{fig:fig1}g shows $C_{sym}$ and $C_{asym}$ at $D=0$. $C_{asym}\approx0$, consistent with inversion symmetry
At low density, $C_{sym}\approx 2c$, so that the change in total electron density matches the geometric capacitance between gates and surfaces.  
This agrees with the expectation from the band theory of Fig. 1a that most of the electronic density of states in this regime is localized on the top and bottom surfaces.  
At high $n_e^0$, in contrast, $C_{sym}\approx 1.8$, indicating a significantly reduced geometric capacitance between carriers and and gates. 
Here surface states are no longer present at the Fermi level, and the electronic density of states resides within the three dimensional bulk as expected from band structure modeling.  Surface and bulk bands are also distinguished by their layer polarizability, $dp/dD$, as shown in Fig. \ref{fig:SI_p_analysis}. 

This picture changes qualitatively at $D=0.72\,\text{V/nm}$, as shown in Fig. \ref{fig:fig1}h.  
Now $C_{asym}$ is nonzero with opposite sign for electron and hole doping consistent with the existence of a single surface state polarized on opposite surfaces for electrons and holes. 
Notably, $C_{sym} \approx 1.93$ is also significantly smaller than 2. This is consistent with the presence of a single surface state whose geometric capacitance to the farther gate is reduced by the series capacitance of the graphite bulk, giving a total geometric capacitance $C_{sym}/c \approx 1+(1+c/c_0)^{-1}$. 
From the comparison of dual- and single surface state regimes, we estimate that $c_0/c\approx 13.3$, implying an out-of-plane dielectric constant  $\epsilon^\perp_{R13G}\approx 16.6$ for the rhombohedral graphite. The enhancement of the rhombohedral graphite dielectric constant as compared to that of hBN can be attributed to the large dielectric response of the  bands away from the Fermi level. 

These findings are summarized in Fig. 1i, 
which shows $C_{sym}$ as a function of $n_e^0$ and $D$. 
The dual-surface regimes manifest as a bright diamond shaped region, with transitions to the single surface regime controlled by a single gate (i.e., the dual surface boundaries occur at constant gate voltage for top- or bottom gates). This single-gate tracking behavior\cite{kolar_single-gate_2025,kumar_superconductivity_2025} can be understood from the electrostatics of flat-band surface states, and arises from the high density of states on the surface which completely screens the potential from the nearer gate. The crossover to single surface polarized regimes at high $|D|$ are confirmed in Fig. 1j, where we show $C_{asym}$  as a function of $n_e^0$ and $D$. In the single surface regime, states at the Fermi level are nearly perfectly layer polarized, manifesting as blue- or red-contrast for bottom- or top-layer polarized carriers, respectively. 

\section{Spin and Valley Ferromagnetism}

In addition to the broad features discussed above, the phase diagrams of Figs. \ref{fig:fig1}i-j also show several sharp features. For several of these, $C_{sym}$ exceeds the maximum expected geometric capacitance of $2c$, indicating negative electronic compressibility of the surface state electron systems\cite{eisenstein_negative_1992}. 
Several additional sharp features emerge in capacitance data at lower temperatures, as shown in Fig. \ref{fig:fig2}a where we plot the penetration field capacitance ($C_{pen}$) at $T\approx 100\,\text{mK}$.  While the sign of features in $C_{pen}$ is difficulty to interpret in multilayer systems, analogies to prior literature on thinner rhombohedral layers suggest these sharp thermodynamic features are associated with the onset of isospin magnetism. 

Electrical transport measurements (Fig. \ref{fig:fig2}b) reveal additional features.  For large values of $D$ and small electron doping, a robust, hysteretic anomalous Hall effect is observed, consistent with valley polarization in a quarter-metal like state.  Notably, anomalous Hall signals persist to at least 8K, as shown in Fig. \ref{fig:fig2}c, considerably higher than the 1-3K reported for R2G and R3G\cite{holleis_fluctuating_2025}.  The high Curie temperature likely arises from the large quasi-degeneracy of the flat band surface states, resulting in a higher density for the quarter metal phases and consequently a higher Coulomb energy.  We also find several superconducting states (Fig. \ref{fig:fig2}d), all of which appear to be pinned close to the phase transitions observed thermodynamically. The superconducting states and their underlying phase transitions follow approximately single-gate tracking trajectories, consistent with a phase transition driven by the physics of a single surface.

To reveal the nature of these transitions, we use tilted-field nanoSQUID on tip microscopy\cite{vasyukov_imaging_2017}. 
nanoSQUID on tip microscopy has been used to directly image magnetism in a variety of flat-band graphitic systems\cite{tschirhart_imaging_2021,auerbach_isospin_2025,arp_intervalley_2024,patterson_superconductivity_2025,uzan_hBN_2025}
To isolate the contribution of the spin magnetic moment, we measure the magnetic field above a sample edge at a physical position chosen to be  sensitive to the fringe fields of in-plane magnetic moments but relatively insensitive to the fringe fields of out-of-plane moment (see Fig. 2e, inset).  Measurements are performed twice--once with magnetic field $(B_\parallel,B_\perp)=(20mT,22mT)$ and once with $(B_\parallel,B_\perp)=(-20mT,20mT)$.  The difference between these two measurements contains information about magnetic moments that change sign with $B_\parallel$, while removing the contributions of purely out-of-plane  orbital moments and other backgrounds which are $B_parallel$-independent\cite{patterson_superconductivity_2025}.  

As shown in Fig. \ref{fig:fig2}e, this measurement modality results in a large signal across both the single- and dual-surface regime. For reference, the fringe magnetic field in the out-of-plane direction at height $h$  above the edge of a semi-infinite sheet of in-plane dipoles is given by $B=\frac{\mu_0 m}{2\pi h}$, giving the observed $500nT$ at $h=180nm$ for $m\approx 5\times 10^{12} \mu_B/cm^{2}$. Given the finite size of the sample, our measurement thus appears consistent with polarization of a substantial fraction of the surface carriers. 
Comparing Figs. \ref{fig:fig2}a,b and e shows that superconductivity appears precisely at the onset of spontaneous spin polarization on the rhombohedral graphene surfaces.  This conclusion is further substantiated using quantum oscillations and the in- and out-of-plane field dependence of the thermodynamic phase transitions, as described in Extended Data Fig. \ref{fig:SI_QO}. 

These findings are summarized in the schematic phase diagram of Fig. \ref{fig:fig2}f.  We contrast this experimental phase diagram with the results of a simplified Hartree-Fock calculation in which we consider isospin symmetry breaking between two flavors.  These simulations account for the band structure but do not attempt to capture details arising from the Hund's coupling or other valley anisotropic interactions (see Methods, as well as  Extended Data Fig. \ref{fig:SI_fock}).  
The experimentally observed boundaries of the spin polarized phase show approximate agreement with where numerical simulation predicts finite isospin polarization. This agreement is tantamount to the statement that the magnetism is driven by the elevated density of states of the flat surface bands.  
One notable quantitative discrepancy is that within Hartree-Fock, symmetry breaking onsets as soon as the conduction band minimum is doped---i.e, at the boundary of the dual-surface region.  Experimentally, however, spin magnetism onsets only at finite doping of one or both  surface bands (see Extended Data Fig. \ref{fig:SI_SPOnset}).  
This discrepancy may indicate the presence of a hidden symmetry-breaking phase between the onset of spin polarization and the surface state band edge.  For example, spin-unpolarized nematics or intervalley coherent  phases would show no contrast in our SQUID measurements, but may be detected by lattice-scale surface probes. 

\section{Surface state superconductivity}

\begin{figure}[ht!]
\centering
\includegraphics[width=\columnwidth]{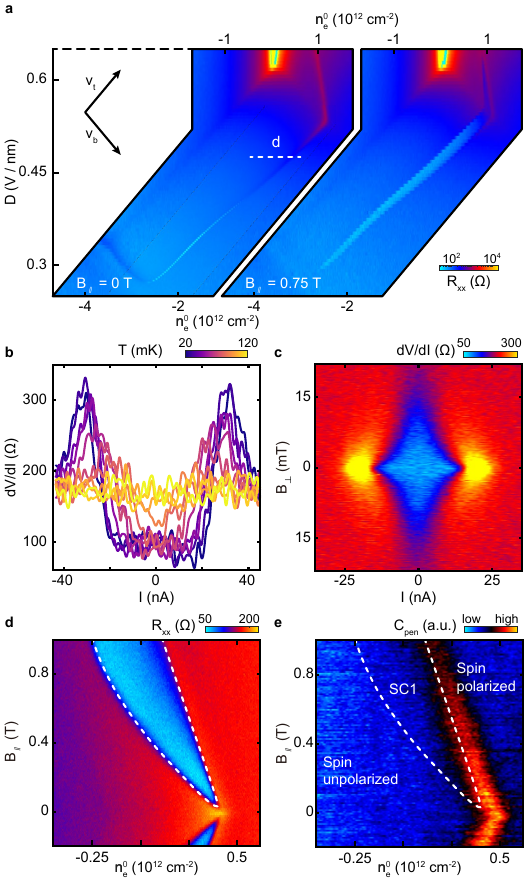}
\caption{
\textbf{Field induced superconductivity and Pauli limit violation in SC1.} 
\textbf{(a)} $R_{xx}$ at B=0 and $B_\parallel=0.75 T$. 
\textbf{(b)} Differential resistance $dV/dI$ at $n_e^0=0.05 \times 10^{12}\,\text{cm}^{-2}$ and $D=0.47 \,\text{V/nm}$ as a function of applied current at different temperatures for $B_\parallel=0.9T$.  $T_C\approx 90mK$ as defined by the onset of the resistivity drop. 
\textbf{(c)} 
$dV/dI$ at $n_e^0=0.30 \times 10^{12}\,\text{cm}^{-2}$ and $D=0.51 \,\text{V/nm}$ as a function of $B_\perp$ at fixed $B_\parallel=0.9 \,\text{T}$.
\textbf{(d)} $R_{xx}$ at small bias as a function of $n_e^0$ and $B_\parallel$ for $D=0.47V/nm$. Superconductivity at this displacement field is induced by a magnetic field.   $T_C$ never exceeds 100mK, implying a Pauli limiting field of $B_p\approx 140mT$ which is exceeded by at least a factor of 7 at the maximum field available. 
\textbf{(e)} Penetration field capacitance as a function of $n_e^0$ and $B_\parallel$ in the same range as panel d.  The bright line is associated with a first order transition to the spin polarized state detected in the magnetization measurements of Fig. 2e. The domain of superconductivity is overlaid and occurs on the spin unpolarized side of the transition. 
}
\label{fig:fig3}
\end{figure}

We now examine the properties of the superconducting state observed on the electron-doped side of the phase diagram (SC1).  At zero magnetic field, SC1 is observed in the dual-surface state regime, tuned primarily by the voltage on a single gate---consistent with superconductivity and the underlying spin transition occurring on a single surface.   
Applying an in-plane magnetic field ($B_{\parallel} = 0.75T$) significantly enhances the domain where superconductivity is observed (Fig. 3a).  Superconducting regions at fixed $D$ become broader in $n_e$, and superconductivity is observed along the same transition lines at values of $D$ where superconductivity is not observed at $B_\parallel=0$.  
Notably, at finite $B_\parallel$, the low resistance state also persists into the single-surface state regime.  
We note that the resistance does not go all the way to zero in our measurements, an observation reported in other works on thicker rhombohedral graphene devices\cite{kumar_superconductivity_2025} as well. 
This may arise from structural defects such as domain walls, which may add a series resistance to an otherwise superconducting sample. We base the claim of superconductivity on several features that have proven robust in other studies of rhombohedral graphene multilayers, namely a sudden onset of lowered resistance at sub-Kelvin temperature accompanied by a strongly nonlinear V-I characteristic and the suppression of these features with small out-of-plane magnetic fields (see Fig. \ref{fig:fig3}b,c and Extended Data Figs. \ref{fig:SI_SC23} and \ref{fig:SI_SC4}).  

Fig. 3d shows the evolution of the low resistance state at $D=0.47$ V/nm, where superconductivity is not observed at zero magnetic field. 
After the onset of superconductivity near $B_\parallel\approx 100mT$, the domain of superconductivity increases with $B_\parallel$ up to our maximum in-plane magnetic field of 1T. Given the low maximum transition temperature $T_c\lesssim 100mK$, SC1 violates the Pauli limit by a factor of at least 5.   This behavior is reminiscent of observations of $B_\parallel$-enhanced or induced superconductivity in rhombohedral graphene bilayers\cite{zhou_isospin_2022} and multilayers\cite{yang_magnetic_2025}.  
The domain of superconductivity in $n_e^0$ also moves with $B_\parallel$, with the high-$n_e^0$ boundary following a linear trajectory. 
Capacitance measurements show that this boundary is associated with the phase transition  between spin unpolarized- and polarized phases observed in magnetometry.  
Comparing Figs. \ref{fig:fig3}d-e shows that superconductivity occurs on the symmetry \textit{unbroken} side of this transition. In other words, the normal state from which SC1 emerges is \textit{not} itself a spin polarized phase, contrasting with, for example, triplet superconductors reported in rhombohedral trilayer\cite{zhou_superconductivity_2021} where the normal state itself is fully spin polarized. The absence of superconductivity at $B_\parallel=0$ also contrasts with superconductors in enhanced spin-orbit graphene multilayers, where singlet- and triplet pairing coexist at $B_\parallel=0$\cite{zhang_enhanced_2023,zhang_twist-programmable_2025,patterson_superconductivity_2025}.  

We propose a simple picture for the dominance of the spin triplet over the spin singlet channel. In the absence of a pairing mechanism that distinguishes between spin flavors, the relative strength of the pairing channels is affected by the Coulomb repulsion.  In rhombohedral graphene multilayers, the ubiquity of spin-polarized half metals rather than, e.g., spin-valley locked half metals implies that Coulomb repulsion is weaker for electrons in opposite valleys with aligned spins than for electrons with opposite spin.  In the language of BCS superconductivity\cite{mcmillan_transition_1968}, then, the effective Coulomb repulsion $\mu^*$ is smaller for triplet pairing. Thus, while our data are of course consistent with an unconventional electronic attraction mechanism that favors the triplet channel specifically, it may suffice that a generalized attractive potential favors triplet pairing owing to the lower Hund's energy for that channel.  

The ferromagnetic and superconducting physics described above are largely properties of a single surface state, and tend to show single gate tracking behavior.  
For these phenomena, the role of the rhombohedral multilayer is to generate a given surface band dispersion, but the opposite surface state (if present at all) plays only a secondary role.  However, some of the broken symmetry states persist to $D=0$, where both surfaces necessarily participate on an equal basis.  Figs. 4a-b show $R_{xx}$ measured in Device 2 near $D=0$ in the vicinity of SC2 and SC3.  
As small $D\approx 0.05V/nm$, two Pauli-limit violating superconducting states (see Extended Data Fig. \ref{fig:SI_SC23}) follow trajectories tuned by opposite gates, indicating they are localized on opposite surfaces.  As $|D|$ is reduced towards zero, the superconducting trajectories appear to `collide' at a low but finite value of $D$, merging into a single superconducting domain in the $n_e^0-D$ plane. This superconducting state remains stable over a finite range of $D$, where it is tuned by the total density rather than the density on the two surfaces individually. 


\begin{figure}[ht!]
\centering
\includegraphics[width=\columnwidth]{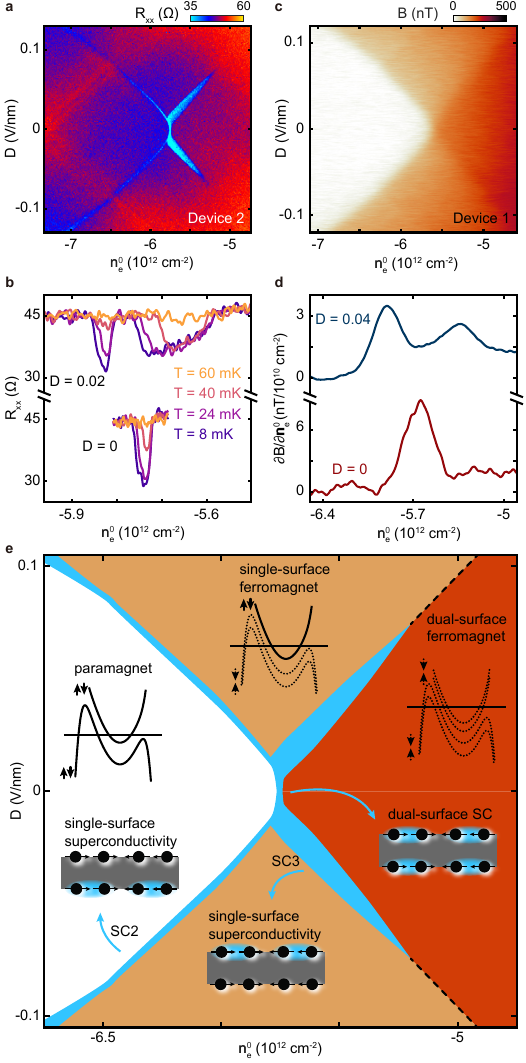}
\caption{\textbf{Dual-surface superconductivity.}
\textbf{(a)} Detail of $R_{xx}$ near SC2 and SC3 at $B_\parallel=0 \,\text{mT}$. 
\textbf{(b)} $R_{xx}$ as a function of temperature at $D=0\,\text{V/nm}$ and $D=0.02\,\text{V/nm}$.  
\textbf{(c)} Magnetometry at $B_{||} = 100$~mT, $B_\perp = 7$~mT, $T = $ 350mK in Device 1, measured over the edge of the device where the fringe field arising from in-plane magnetic moments dominates that arising from out of plane moments.   
\textbf{(d)} Derivative of measured magnetic field with respect to electron density,  at $D=0$ and $0.04 \,\text{V/nm}$.
\textbf{(e)} Schematic phase diagram showing regions of single- and dual-surface spin ferromagnetism as well as single- and dual-surface superconductivity. 
}
\label{fig:fig4}
\end{figure}


Measurements of the fringe magnetic field $B$ in the same regime of parameter space for Device 1 are shown in Figs. 4c-d (see also additional data in Extended Data Fig \ref{fig:SI_SC23}).  The two superconductors apparently follow steps in the spin magnetization. At finite $D$, these steps correspond to independent steps in the spin magnetization on the two surfaces.  We can thus associate the two superconductors with distinct spin-polarization transitions occurring on opposite surfaces.  
Near $D=0$, a single step is observed, corresponding to the simultaneous onset of spin magnetization on both surfaces.  By symmetry, superconductivity at $D=0$ should be equally distributed on both surfaces. 

It is interesting to consider the implications of $D=0$ superconductivity in a system with such strongly surface polarized wavefunctions.  At the single particle level, the band index is a good quantum number of the clean system. However, the conduction and valence bands are nearly degenerate at $D=0$.  Electronic interactions (whether direct or mediated by a neutral phonon or magnetic modes) will generically be stronger within a surface than between surfaces.  Superconducting pairing is thus not likely to be layer indifferent, and the  $D=0$ superconducting  system should be understood as two superconductors on opposite surfaces, Josephson coupled via the fully insulating graphite bulk.  The properties of such single crystal Josephson junctions will be an interesting avenue for future work.  If surface-selective contacts can be developed, the Josephson energy could be determined by direct current measurements of the inter-surface critical current.  Alternatively, finite frequency interlayer capacitance measurements of the type presented here might detect the low frequency Josephson plasmon, which constitutes an extreme limit of infrared active collective modes proposed to exist in all rhombohedral graphene superconductors\cite{levitan_linear_2024}.

\section{Discussion}

Our results show that the correlated electron physics of RNG persist to large layer number, where the D-tunable two dimensional electron systems for small-$N$ RNG evolve into the physics of separate surface states at the boundaries of the insulating, incompressible three dimensional graphite bulk for large $N$.  In several key respects, thick rhombohedral graphite shows similar phenomenology to graphene bilayers and other rhombohedral multilayers, with competing half- and quarter metals, ferromagnetic Hund's coupling, and numerous low temperature superconducting phases appearing near phase boundaries.  As shown in Fig. \ref{fig:SI_R14G}, alignment of thick rhombohedral graphite to the hBN can also induce symmetry breaking insulating phases, again analogous to thinner layers. One notable quantitative advantage is the maximum density that can be induced without populating dispersive bulk states---i.e., the quasi-degeneracy of the flat band.  Flat band physics in all graphene multilayers, including both rhombohedral and rotationally faulted stacking orders, arises from the interplay of intralayer hopping (parameterized by the monolayer graphene Fermi velocity $v_f\approx 10^6 \,\text{m/s}$) and interlayer hopping, which is dominated by the dimer coupling $\gamma_1\approx 400 \,\text{meV}$.   The scale of density over which flat bands can be induced is then set by $\Delta n_e\approx (\gamma_1/ \hbar v_f)^2/\pi\approx 1.2\times 10^{13}\,\text{cm}^{-2}$. Few-layer graphene systems typically fall well short of this value.  As is evident in our data, however, physical R13G comes close this limit, with surface polarized quarter metal phases observed at carrier densities as high as $n_e\approx3\times 10^{12}\,\text{cm}^{-2}$, implying $\Delta n\approx 1.2 \times 10^{13} \,\text{cm}^{-2}$ for the conduction band minimum.  Indeed, the high temperature for the anomalous Hall onset at 8K is nearly an order of magnitude higher than in crystalline bilayer graphene, evidence for the larger scale of Coulomb repulsion. This near-order-of-magnitude enhancement in the critical temperature of the correlated states opens the possibility of bringing optical techniques to bear on the correlated properties of these multilayer systems.

Despite the increased scale of the Coulomb repulsion, the superconducting transition temperatures observed here remain comparable to superconductors in other few-layer rhombohedral graphene systems. This apparent decoupling of the superconducting and magnetic critical temperatures as a function of $N$ suggests the possibility of using the layer number $N$ as a rhombohedral graphene analog of the ``isotope effect'', constraining theoretical models for the pairing mechanism, and may prove a fruitful avenue for future theoretical and experimental work.

\section{Acknowledgments}
 The authors would like to acknowledge discussions with E. Berg, Y. Vituri, A. H. Macdonald, Vo Tien Phong, A. Berenevig, and J. Herzog-Arbeitman. 
Work in the Young lab at UCSB was primarily supported by the Department of Energy under Award DE-SC0020043 to A.F.Y., and by the Gordon and Betty Moore Foundation, grant DOI 10.37807/GBMF13801.
Numerical calculations are done using the High Performance Compute Cluster of the Research Computing Center (RCC) at Florida State University. K.K. was supported by a grant from the Simons Foundation (SFI-MPS-NFS-00006741-12, P.T.) in the Simons Collaboration on New Frontiers in Superconductivity.  
C.L. was supported by start-up funds from Florida State University and the National High Magnetic Field Laboratory. 
The National High Magnetic Field Laboratory is supported by the National Science Foundation through NSF/DMR-2128556 and the State of Florida. 
K.W. and T.T. acknowledge support from the Elemental Strategy Initiative conducted by the MEXT, Japan (Grant Number JPMXP0112101001) and JSPSKAKENHI (Grant Numbers 19H05790, 20H00354 and 21H05233). 
A portion of this work was performed in the UCSB Nanofabrication Facility, an open access laboratory.  
T.A. and O.S. acknowledge direct support by the National Science Foundation through Enabling Quantum Leap: Convergent Accelerated Discovery Foundries for Quantum Materials Science, Engineering and Information (Q-AMASE-i) award number DMR-1906325; the work also made use of shared equipment sponsored under this award. 

\clearpage
\newpage
\pagebreak

\onecolumngrid

\section{Methods}

\subsection{Sample preparation}
Graphene and hexagonal boron nitride (hBN) flakes were prepared by mechanical exfoliation onto $\mathrm{SiO_2/Si}$ substrates. The rhombohedral stacking order of the graphene was identified using a combination of Raman spectroscopy\cite{lui_imaging_2011,zhang_raman_2016}, infrared (IR) imaging\cite{lu_extended_2025,feng_rapid_2025}, and Scanning Microwave Impedance Microscopy (SMIM) \cite{holleis_nanoscale_2025} (see Extended Data Fig. \ref{fig:SI_device}). To prevent relaxation during subsequent processing, the rhombohedral domain was isolated by Atomic Force Microscopy (AFM) anodic oxidation\cite{li_electrode-free_2018,zhou_half-_2021}. 
To ensure symmetric device geometry, special care was taken in dielectric preparation. For devices 1 and 2, the top and bottom hBN dielectrics were prepared by cleaving a single, thin flake into two identical halves using a (P-50 PtIr) STM tip on the transfer station.

The heterostructure stacks were assembled in three steps. First, the bottom hBN dielectric and the graphite gate, were sequentially picked up with a poly(bisphenol A carbonate) (PC) film and released onto the $\mathrm{SiO_2}$ substrate. After dissolving the PC film, the surface of the bottom dielectric was cleaned with a AFM tip in contact-mode to remove polymer residues. Next, the top hBN dielectric and the rhombohedral graphene were similarly picked up and released onto the pre-cleaned bottom stack. A second AFM tip cleaning step was performed on the top surface, and the preserved rhombohedral stacking order was verified using Raman spectroscopy and IR imaging. A final graphite top gate, exfoliated on a PDMS stamp, was then transferred on top to complete the stack. This stack was fabricated into the measurement device using standard e-beam lithography, reactive-ion etching and e-beam deposition of metal contacts.

\subsection{Capacitance measurements}
To measure the capacitance of the device, which is typically several orders of magnitude smaller than the parasitic capacitance inherent in the measurement setup, we utilize a custom AC bridge circuit and high electron mobility transistors (HEMTs) as amplifiers inside the cryostat. The schematic of the circuit is shown in Fig. \ref{fig:SI_CapCircuit}. 

The rhombohedral graphene and bottom gate are connected to one branch of the bridge circuit, in parallel with a reference capacitor of constant value. Two AC voltages with tunable relative phase and amplitude are applied across the sample and the reference capacitor. The output is nulled out at the balance point. Subsequent deviation caused by changes in sample capacitance, are monitored by applying the signal to the gate of a reading HEMT (HEMT1 and HEMT4), which converts the sample impedance of femtofarads into $\SI{1}{\kohm}$ output impedance for subsequent readout at room temperature. 

To enable the simultaneous, quantitative measurement of $C_{t/b}$ and $C_{pen}$ in-situ, two additional HEMTs (HEMT2 and HEMT3) are integrated. These allow us to tune the impedance of each measurement branch while maintaining the amplifier input at high impedance and performing capacitance readouts from the opposing branch.
For instance, when measuring $C_b = \frac{\partial (n_1+n_2)}{\partial v_b}$ and $C_t = \frac{\partial (n_1+n_2)}{\partial v_t}$, HEMT2 is fully depleted ($V_2=-0.5 \,\text{V}$) and HEMT3 bypasses the $\SI{100}{\Mohm}$ resistor ($V3=0\,\text{V}$), so that the AC signal $v_b$ is applied to the bottom gate through a low-impedance path while maintaining the high impedance of the measurement transistor (HEMT1) input. To measure $C_{pen}=\frac{\partial n_b}{\partial v_t}$ and $C_b=\frac{\partial n_b}{\partial v_s}$, the configuration is reversed, such that HEMT3 is depleted and HEMT2 is a short circuit to provide a low-impedance path for $v_s$. 
Since we utilize a bridge circuit, the measured capacitance quantities are relative to the reference capacitors $c_{ref1/2}$ of unknown but constant values. To determine the ratio of these reference capacitors $c_{ref1}/c_{ref2}$, we employ the physical symmetry constraint, $C_b\equiv \frac{\partial n_b}{\partial v_s}=\frac{\partial (n_1+n_2)}{\partial v_b}$, by measuring the same quantity relative to both $c_{ref1}$ and $c_{ref2}$. This allows $C{pen}$ to be quantitatively compared in the same units as $C_{t/b}$. Measurements of these three calibrated capacitance quantities can be used to derive charge carrier density $n$ and layer polarization $p$, as discussed in the next section. 
The layer-resolved capacitance measurement was performed in a Helium-4 cryostat. Signals at 68.177kHz were generated by a custom-built dual AD9854 "AC box", which provides two phase-locked output channels with tunable phase and amplitude. A Stanford Research Systems 860 lock-in amplifier was used to readout the output signals. The parasitic capacitances of the measurement setup $C_{par}/c$ have been subtracted, 0.56 for $C_{pen}$, 0.60 for $C_t$ and 1.69 for $C_b$. These constant offsets are determined by measuring $C_{pen}$ within the ''diamond'' (dual-surface region) and $C_{t/b}$ inside the gap, where both values are expected to be zero. The value of $c$ is given by $C_{sym}=2c$ in the dual-surface region.

\subsection{Four-plate capacitor model}
To interpret the capacitance data and derive an approximate value for the layer polarization, we model the dual-gated rhombohedral graphite as a four-plate capacitor, where electrons are assumed to reside only on the outer layers, separated by the dielectric bulk $c_0$. $c_t=c_b=c$ denotes the geometric capacitance between the sample and the gates, and $n_i$ corresponds to $electron$ densities on each plate, as illustrated in Fig \ref{fig:SI_CapCircuit}.

Starting from electrostatics and conservation of charge, we get the following four equations.

\begin{align}
  n_t + n_b + n_1 + n_2&= 0\\ 
  -n_t &= c (v_t - \phi_1)\\
  -\frac{n_t + n_1 - n_b -n_2}{2} &= c_0 (\phi_1 - \phi_2)\\
  -n_b&= c (v_b - \phi_2) 
\end{align}

After getting rid of $\phi_1$ and $\phi_2$, we take the derivatives of the two remaining equations with respect to $v_t$ and $v_b$, and solve for $\frac{\partial n_{1(2)}}{\partial v_{t(b)}}$:
\begin{align}
    \frac{\partial n_1}{\partial v_t} &= C_p+C_t+\frac{c_0(-c+2C_p+C_t)}{c}\\
    \frac{\partial n_1}{\partial v_b} &= c_0-C_p-\frac{c_0(C_b+2C_p)}{c}\\
    \frac{\partial n_2}{\partial v_t} &= c_0-C_p-\frac{c_0(C_t+2C_p)}{c}\\
    \frac{\partial n_2}{\partial v_b} &= C_p+C_b+\frac{c_0(-c+2C_p+C_b)}{c}
\end{align}
where $C_t= \frac{\partial (n_1+n_2)}{\partial v_t},\, C_b= \frac{\partial (n_1+n_2)}{\partial v_b},~ \text{and} ~ C_p= \frac{\partial n_t}{\partial v_b} = \frac{\partial n_b}{\partial v_t}$ are the measured quantities in the experiment.
The derivatives of the total density ($n \equiv  n_1+n_2$) and layer polarization ($p\equiv n_1-n_2$) with respect to the density ($n_0\equiv c(v_t+v_b)$) and the displacement field  ($p_0\equiv c(v_t-v_b)=2\epsilon_0D$) induced by the applied voltages can be rewrite as the followings:
\begin{align}
    \frac{\partial n}{\partial n_0} &= \frac{\partial n}{\partial v_t} \frac{\partial v_t}{\partial n_0} + \frac{\partial n}{\partial v_b} \frac{\partial v_b}{\partial n_0} = \frac{C_t}{2c} + \frac{C_b}{2c} = \frac{C_{sym}}{2c} \label{eq:dndn0}\\
    \frac{\partial n}{\partial p_0} &= \frac{\partial n}{\partial v_t} \frac{\partial v_t}{\partial p_0} + \frac{\partial n}{\partial v_b} \frac{\partial v_b}{\partial p_0} = \frac{C_t}{2c} - \frac{C_b}{2c} = \frac{C_{asym}}{2c} \label{eq:dndp0}\\
    \frac{\partial p}{\partial p_0} &= \frac{\partial p}{\partial v_t} \frac{\partial v_t}{\partial p_0} + \frac{\partial p}{\partial v_b} \frac{\partial v_b}{\partial p_0} \nonumber\\
    \frac{\partial p}{\partial n_0} &= \frac{\partial p}{\partial v_t} \frac{\partial v_t}{\partial n_0} + \frac{\partial p}{\partial v_b} \frac{\partial v_b}{\partial n_0} \nonumber\\
    &=\frac{1}{2c} \left( \frac{\partial n_1}{\partial v_t} + \frac{\partial n_1}{\partial v_b} - \frac{\partial n_2}{\partial v_t} - \frac{\partial n_2}{\partial v_b}\right) \nonumber\\
    &=(\frac{1}{2}+\frac{c_0}{c})\frac{C_{asym}}{c} \label{eq:dpdn0}\\
    &=\frac{1}{2c} \left( \frac{\partial n_1}{\partial v_t} - \frac{\partial n_1}{\partial v_b} - \frac{\partial n_2}{\partial v_t} + \frac{\partial n_2}{\partial v_b}\right) \nonumber\\
    &=\frac{c_0}{c}\frac{4C_p+C_{sym}-2c}{c}+\frac{4C_p+C_{sym}}{2c} \label{eq:dpdp0} 
\end{align}
where $C_{sym/asym} = C_t \pm C_b$ as defined in the main text.

\subsection{NanoSQUID on Tip measurements}

The nanoSQUID on Tip (nSOT) measurements were performed using an indium SQUID fabricated on the end of a pulled quartz pipette using the self-aligned fabrication technique  with a diameter of about 312~nm and a sensitivity of $\approx$ 4~nT/$\sqrt{\textrm{Hz}}$. Readout was completed using a series SQUID array amplifier in feedback mode. The tip was angled 12$^{\circ}$ off of vertical to enable flux biasing with magnetic field in the plane of the sample.
All nSOT measurements were performed in a Helium-3 cryostat with the sample in vacuum, at the base temperature of 340~mK.
All magnetometry measurements were completed on Device 1.

The nSOT measurements were taken using a quasi-DC contrast mechanism consisting of modulating the gates in a square wave between two voltages at a frequency typically between 1 and 2 kHz using Analog Devices DG417 analog switch. 
We use 50kHz single pole RC filters to smooth the waveform slightly. 
In Figure 2e and 4c the ``reference" gate value was $n_e = 1.554 \times 10^{12} \,\, \mathrm{cm}^{-2}$ and $D = 0$. 
This ``boxcar" data is then read out using an Stanford Research Systems 830 lock-in amplifier to extract the amplitude of modulation between the point of interest and the non-magnetic reference point, i.e. the relative change in fringe magnetic field. 
In practice, a lock-in amplifier is designed to quantitatively measure a sine waveform, not a square waveform and therefore a small scaling factor, determined by quantitatively comparing a lock-in voltage measurement to a known size input square voltage waveform.
The scaling factor for our analog switches and filtering is $F_\text{scale} = 0.447$ and is applied to the data uniformly such that $B_\text{reported} = B_\text{lock-in}/F_\text{scale}$.
We independently check that our scaling is correct for each measurement by directly measuring the difference between the two levels of the boxcar waveform using the Zurich Instruments MFLI MF-BOX Boxcar Averager.

\subsection{Theoretical modeling}

\subsection{Electrostatic description of the low-energy rhombohedral multilayer graphene}

We model rhombohedral graphene by considering the following single-particle Hamiltonian
\cite{mccann_landau-level_2006,nilsson_electronic_2008,mccann_electronic_2013}
\begin{equation}
\hat H_{SP}=
\label{eq:hambernal}
\sum_{\mathbf k}
\Psi^\dagger_{\mathbf k}
    \begin{pmatrix}
        U_1 &  v_F \overline{k}&-v_4 \overline{k}&-v_3 k &-t_2/2&0& \\
         v_F k& U_1&t_1  &-v_4 \overline{k} &\\ 
         -v_4 k& t_1&U_2 &  v_F \overline{k}& \\
        -v_3  \overline{k}&-v_4 k& v_F k &U_2&  \\
        -t_2/2& &&& \ddots\\
        \vdots&  
    \end{pmatrix}
\Psi_{\mathbf k} 
\end{equation}
in the basis of sublattices ($A$, $B$) and layers ($1,2,\dots$) - $\Psi_{\mathbf k} = (c_{\mathbf k,1,A}, c_{\mathbf k,1,B}, c_{\mathbf k,2,A}, c_{\mathbf k,2,B}, \dots)$. Here $k=k_x+ik_y$, $\overline{k}=k_x - ik_y$ momenta are measured from the $\vec{K}$ point, $v_F$ is the graphene Dirac velocity,
and $v_3, v_4 \ll v_F$ denote nonlocal interlayer tunneling velocities, and $t_1$ is the strength of $B1\to A2$ tunneling.
Crucially, $U_l$ is the layer-dependent electrostatic potential energy.
We use the parameters from \cite {park_topological_2023}. Specifically, we take
$v_F = \SI{-547}{meV\cdot nm}$, $v_3=\SI{61.66}{meV\cdot nm}$ and
$v_4  = \SI{30.3}{meV\cdot nm}$, $t_1   =  \SI{356.1}{meV}$
$t_2  =  \SI{-8.3}{meV}$. In our analysis we do not include additional potentials on the outer layers as we determine the layer potential fully self-consistently by electrostatics as detailed below.

A key aspect of our analysis relies on a self-consistently determined layer-dependent electrostatic potentials  $U_l$. A solution for these potentials is given by Gauss' law as (see also Refs.\cite{kolar_single-gate_2025, kolar_electrostatic_2023})
\begin{equation}
\label{eq:potdifference}
    U_{l+1}-U_{l} = -e^2 d \frac{\dens_b+ \sum_{j \leq l}\dens_j }{\epsilon_\perp\epsilon_0}.
\end{equation}
with $d$ being the interlayer distance and $\epsilon_\perp$ corresponding to the out-of-plane dielectric constant. 
We denote the net electron densities in each layer
$l$ by $\dens_l$, and in the top and bottom gates by $\dens_t$ and $\dens_b$,
respectively. Overall charge neutrality of the whole system (sample and the gates) implies that
the sum of these gate charges fixes the total device density, $\dens = \sum_l
\dens_l = -(\dens_t + \dens_b)$. Their difference sets the experimentally
accessible displacement field, $D= e\frac{\dens_b-\dens_t}{2\epsilon_0}$

At a given combination of $\dens_t$ and $\dens_b$, we need to solve Eq.~\eqref{eq:hambernal}
together with Eq.~\eqref{eq:potdifference} self-consistently.
The solution is a density matrix
$[P(\mathbf k)]_{l,s, l',s'} = \langle 
c^\dagger_{\mathbf k,l,s }c^{\phantom{\dagger}}_{\mathbf k,l',s'}
\rangle$,
which determines the layer densities as 
$\rho_l =\frac{N_{\text{deg.}}}{A}\sum_{\mathbf k,s}  \left\{[P(\mathbf k)]_{l,s, l,s}- \frac{1}{2} \right\},$
where the factor of $1/2$ subtracts the density at charge 
neutrality
\cite{kwan_moire_2025}.

Since the above Hamiltonian corresponds to the K-valley only, and as we do not consider intervalley coherent states, 
we model a symmetric state with fourfold degeneracy by multiplying the density of the K-valley state by $N_{\text{deg.}}=4$.

\subsection{Coulomb-driven symmetry-breaking}

To study interaction effects beyond layer potentials, and states without perfect flavor degeneracy (symmetric states), 
we consider $N_{flavor}$ copies of the above single-particle Hamiltonian, modelling $N_{flavor}$ flavors as
\begin{equation}
\hat H = \sum^{N_{flavor}}_f \hat H_{SP,f} +H_{\text{int}},
\end{equation}
adding the following interaction term
\begin{equation}
H_{\text{int}} = \frac{1}{2A} \sum_{\mathbf{q}\neq 0}\sum_{l,l'} V_{ll'}(\mathbf q) :\rho_{\mathbf q,l} \rho_{-\mathbf q,l'}:,
\label{eq:interactinghamiltonian},
\end{equation}
where $V_{ll'}(\mathbf q)$ is the double-gate screened Coulomb interaction \cite{kolar_electrostatic_2023}. Here $A$ is the system area, $::$ denotes normal ordering, and $\hat \rho_{-\mathbf q,l}$ is the charge density operator in layer $j$ at momentum $\mathbf q$.
The double-gate screened Coulomb interaction reads \cite{kolar_electrostatic_2023}
\begin{align} 
\label{eq:doublegatecoulomb}
 V_{ll'}(\mathbf q)  =
\frac{1}{2\epsilon \epsilon_0}\frac{1}{q}\cdot \left(\frac{e^{-q (z_l+z_{l'})} 
\left(-e^{2 q (d_s+z_l+z_{l'})}-e^{2 d_s q}+e^{2 q z_l}+e^{2 q z_{l'}}\right)}{e^{4 d_s q}-1}+e^{-q |z_l-z_{l'}|} \right).
\end{align} 
where $\epsilon$ is the dielectric constant, which we allow to differ from $\epsilon_\perp$ as it accounts primarily for in-plane interactions, 
and where $d_s$ is the gate-sample distance.

Within our approach, we first determine a self-consistent density matrix 
$P(\mathbf k)$ from the symmetric calculation, which includes only the layer potentials.
We project on $N_{\text{active}}$ bands of this calculation closest in energy to the Fermi level,
with wavefunctions $\ket{u_{\mathbf k, \alpha}}$, $\alpha=1,\ldots, N_{\text{active}}$, where the operator
for each band for flavor $f$ is $d^\dagger_{\mathbf k,f,\alpha }$.
Transforming from the band to orbital basis is done using the matrix
$U_{ls, \alpha}(\mathbf k) = \braket{ls| u_{\mathbf k, \alpha}}$, where $l$ is a layer index and $s$ a sublattice index,
appropriate for rhombohedral $N$-layer graphene.
The projector operator on the active bands is
$\proj(\mathbf k) = \sum_{\alpha=1}^{\nflav}\ketbra{u_{\mathbf k, \alpha}}$,
 which can be expressed in orbital basis as
 \begin{equation}
\proj(\mathbf k)|_{ls,l's'} = 
\sum_{\alpha=1}^{\nflav} U_{ls, \alpha}(\mathbf k)
U^\dagger_{\alpha,l's'}(\mathbf k)\,.
 \end{equation}

We assume that all the bands below the $\nactive$ bands are fully filled and denote by $P^0(\mathbf k)$ the density matrix (in layer and sublattice basis) of the filled remote bands.
It can be obtained as  
\begin{equation}
P^0(\mathbf k) =   P(\mathbf k) - \proj(\mathbf k) P(\mathbf k) \proj(\mathbf k).
\end{equation}
The filled remote bands contribute an additional Fock term, which we need to include and present below. 

Written in terms of the density matrix
$[P_f(\mathbf k)]_{\alpha \beta} = \langle 
d^\dagger_{\mathbf k,f,\alpha }d^{\phantom{\dagger}}_{\mathbf k,f,\beta}  
\rangle$,
the mean-field Fock term reads
\begin{align}
 \label{eq:hmffock}
\hat H_{Fock} &=-\frac{1}{A}\sum_{f}\sum_{l,l'}\sum_{\mathbf q,\mathbf k}  
V_{l,l'}(\mathbf q) 
\nonumber \\ 
&\times d^\dagger_{\mathbf k,f}\left[
\Lambda^{l}_{\mathbf q}(\mathbf k) P^T_f(\mathbf k + \mathbf q) \Lambda^{l'}_{- \mathbf q}(\mathbf k +\mathbf q)
\right]d_{\mathbf k,f},
\end{align}
where $\Lambda^{l}_{\mathbf q}$ is the overlap (form factor) matrix:
\begin{equation}
[\Lambda^{l}_{\mathbf q}(\mathbf k)]_{\alpha,\beta} =\braket{u_{\mathbf k, \alpha}|u_{\mathbf k+\mathbf q, \beta}}.
\end{equation}

The remote Fock term due to the filled valence bands, whilst subtracting a background. Explicitly, it reads
\begin{align}
\hat H^{Remote}_{Fock} &=-\frac{1}{A}\sum_{f}\sum_{l,l'}\sum_{\mathbf q,\mathbf k}  
V_{l,l'}(\mathbf q) 
\nonumber \\ 
&\times d^\dagger_{\mathbf k,f} U^\dagger(\mathbf k)
\left\{(P^0)^T(\mathbf k + \mathbf q) - \frac{1}{2}\delta_{l,l'}\delta_{s,s'}
\right\}U(\mathbf k) d_{\mathbf k,f}.
\end{align}

The last component of the mean-field analysis is an expression for the layer densities, which are in this projected approach given as
\begin{equation}
\rho_l =\frac{1}{A }\sum_{\mathbf k}  \left\{
\frac{\ndeg}{\nflav} \sum_f \mathrm{Tr}\left[ \Lambda^{l}_{\mathbf q=0}(\mathbf k) P^T_f(\mathbf k ) \right]+
\ndeg \sum_{s}\left( [P^0(\mathbf k)]_{ls, ls}- \frac{1}{2}\right) \right\},
    \end{equation}
    where the first term is the density of the active bands, while the second term is the remote charge
density due to the filled valence bands, with the density matrix at infinite temperature subtracted.

To summarize, the mean-field Hartree-Fock Hamiltonian reads
\begin{equation}
\hat H_{\text{MF}} = \sum^{N_{flavor}}_f \hat H_{SP,f} + \hat H^{Remote}_{Fock} + \hat H_{Fock}.
\end{equation}

\subsection{Details of the numerical implementation}
We use the optimal damping algorithm \cite{cances_can_2000} to aid convergence in the self-consistent calculation. For the pure electrostatic calculations, we assume a $\cth$ symmetric solution. This $\cth$ solution assumption allows the use of a reduced $\mathbf k$ grid, in which points related by $\cth$ symmetry are identified.
To generate the band structure plots, we use a grid of 1449 $\mathbf k$-points centered around $\mathbf k=0$ with radius
$|\mathbf k_{max}| = \SI{1.366}{nm^{-1}}$ 
In our calculation, we use an out-of-plane dielectric constant of $\epsilon_\perp=3$ expected for graphite. This value of the dielectric constant is appropriate, as we treat the screening of layer potentials by remote bands consistently. 

For the Fock calculations, we simulate $\nflav=2$ spin flavors, thus
assuming a twofold $\ndeg=2$ valley degeneracy. 
Unlike in the pure electrostatic calculations, we do not assume $\cth$ symmetry, and our $\mathbf k$ grid has 379 $\mathbf k$ points
with radius
$|\mathbf k_{max}| = \SI{0.928}{nm^{-1}}$ 
Here, we use a larger in-plane dielectric constant, $\epsilon=10$, to account for remote-band screening. 
At each point in the $n-D$ plane, we find the optimal state by proposing two candidate states -- the layer antiferromagnet
and the ferromagnet. We let the iterations converge for each initial state and obtain the final phase diagram by choosing the 
lower energy solution.

Finally, we note that the maximal extent of the flat band is $|\mathbf k_c| =\frac{t_1}{v_F} = \SI{0.65}{nm^{-1}}$. Thus, both momentum cut-offs (for both Hartree and Fock calculations) used for simulations include the entire flat band.

\newpage
\bibliography{references}

\clearpage
\setcounter{figure}{0}
\renewcommand{\thefigure}{S\arabic{figure}}


\begin{figure*}[t]
\centering
\includegraphics[width=\textwidth]{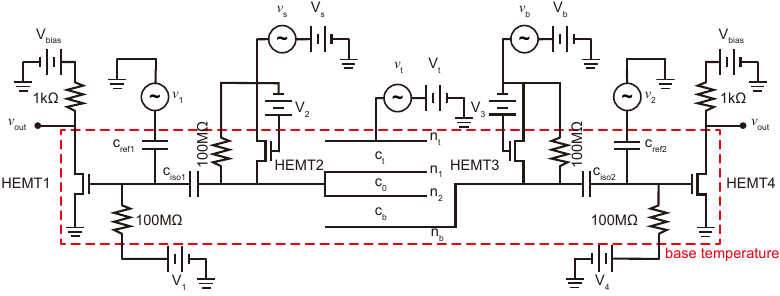}
\caption{
\textbf{Detailed circuit schematic for capacitance measurements.}
DC voltages are denoted $\text{V}_i$ while AC voltages are denoted $v_i$. $\text{V}_{1-4}$ are used to tune the impedance of the HEMTs (model FHX35X). $\text{V}_{t/b/s}$ are applied in combination to control $n_e^0$ and $D$ within the device. AC excitations $v_{t/b/s}$ are applied to measure the relevant capacitance components. The AC signals are first nulled by balancing them with $v_{1/2}$ through the reference capacitors $c_{ref1/2}$. The resulting output signal from sweeping $n_e$ and $D$ is read out at room temperature as a voltage ($v_{\text{out}}$).
}
\label{fig:SI_CapCircuit}
\end{figure*}

\clearpage


\begin{figure*}[t]
\centering
\includegraphics[width=\textwidth]{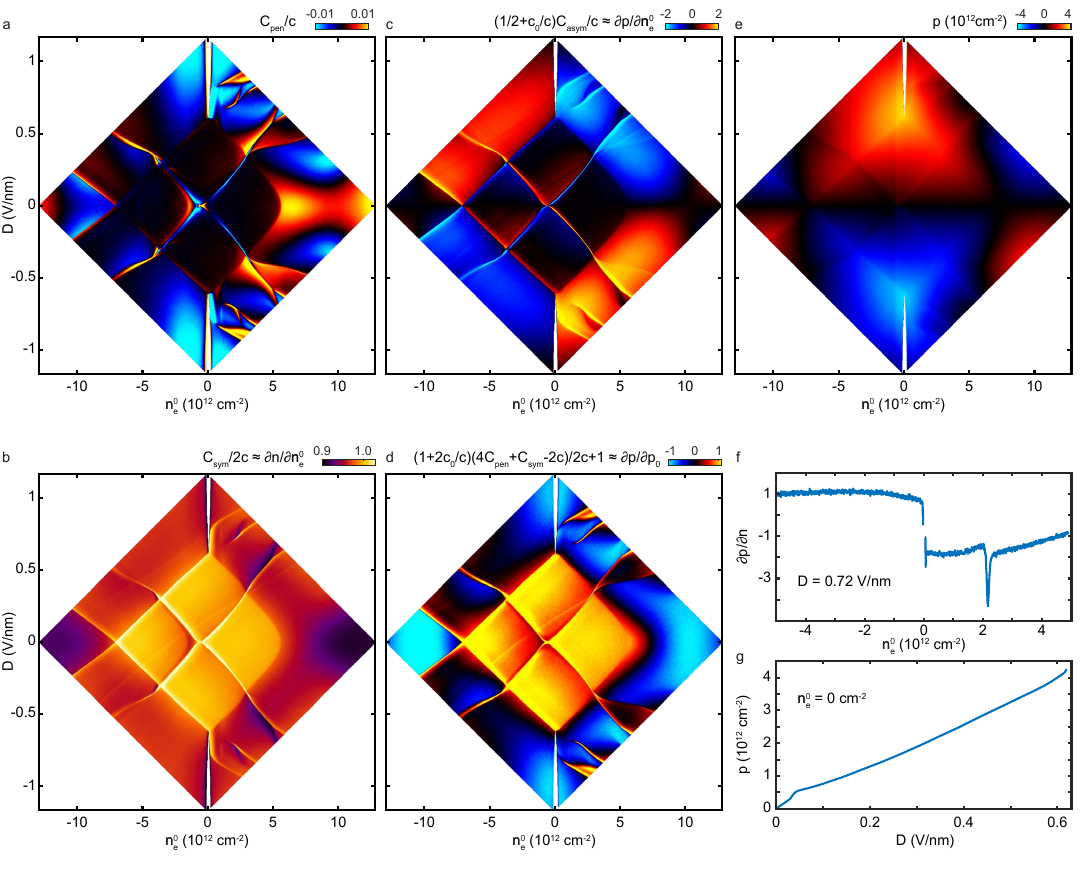}
\caption{
\textbf{Penetration capacitance $C_{pen}$ and layer polarization $p$.} 
\textbf{(a)} $C_{pen}$ normalized by $c$. 
\textbf{(b-d)} Derived differential quantities calculated from the experimental dataset of $C_{pen}$, $C_t$, and $C_b$, following equations \ref{eq:dndn0},\ref{eq:dpdn0} and\ref{eq:dpdp0}: \textbf{(b)}$\partial n/\partial n_e^0$, \textbf{(c)}$\partial p/\partial n_e^0$, and \textbf{(d)} $\partial p/\partial p_0$ .
\textbf{(e)} Layer polarization $p$ integrated from the $dp/dp_0$ data shown in panel d.
\textbf{(f)} $(\partial p/\partial n)_{p_0} = \frac{\partial p/\partial n_e^0}{\partial n/\partial n_e^0}$ at $D=0.72 \,\text{V/nm}$
\textbf{(g)} $p(D)$ at charge neutrality.
}
\label{fig:SI_p_analysis}
\end{figure*}

\clearpage


\begin{figure*}[t]
\centering
\includegraphics[width=\textwidth]{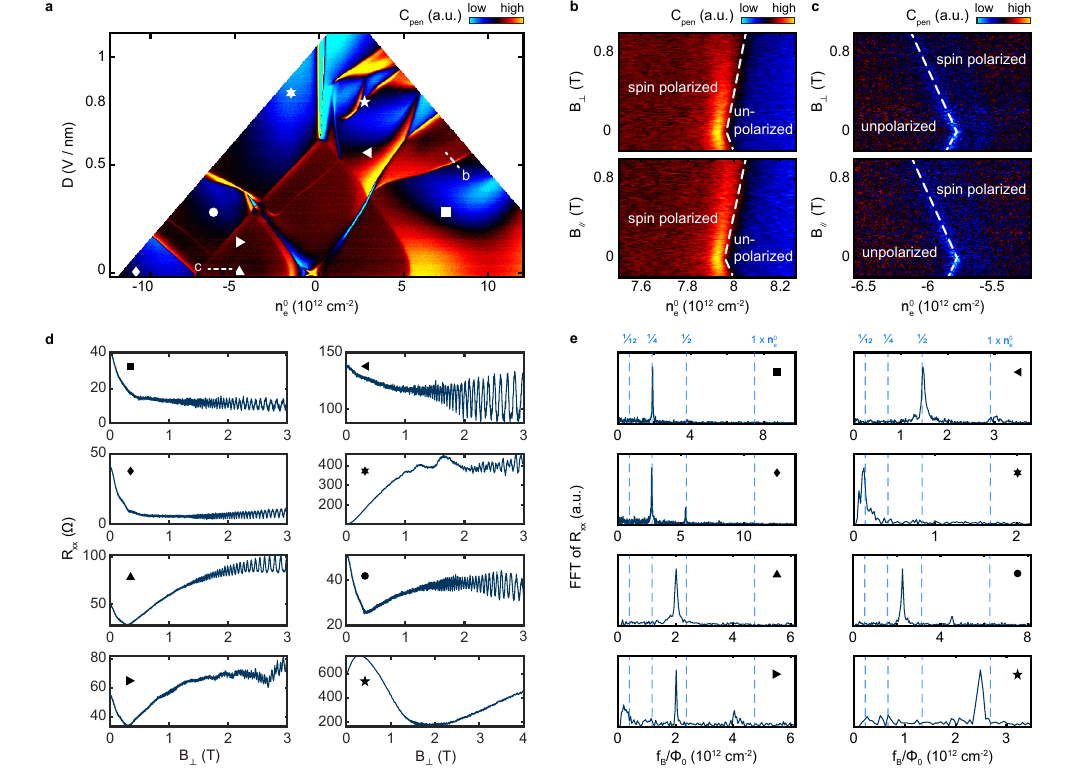}
\caption{
\textbf{Phase transition magnetic susceptibility and quantum oscillations.} 
\textbf{(a)} $C_{pen}$ map reproduced from Fig. 2a. 
\textbf{(b)} and \textbf{(c)} Magnetic field dependent $C_{pen}$ across the phase transitions between spin unpolarized and polarized phases on electron- and hole- doped side, acquired at along the trajectories indicated in panel a.  Spin polarized phases are thermodynamically favored over unpolarized phases for both in- and out-of-plane fields. 
\textbf{(d)} Quantum oscillations measured at the $n_e-D$ point indicated in panel a.  
\textbf{(e)} Fourier transform of $R_{xx}(1/B_\perp)$ for the curves shown in panel d. The blue dashed lines indicate fractions of $n_e^0$, serving as a visual guide to demonstrate the different degeneracy of the Fermi surface across the whole phase diagram.
}
\label{fig:SI_QO}
\end{figure*}

\clearpage

\begin{figure*}[t]
\centering
\includegraphics[width=\textwidth]{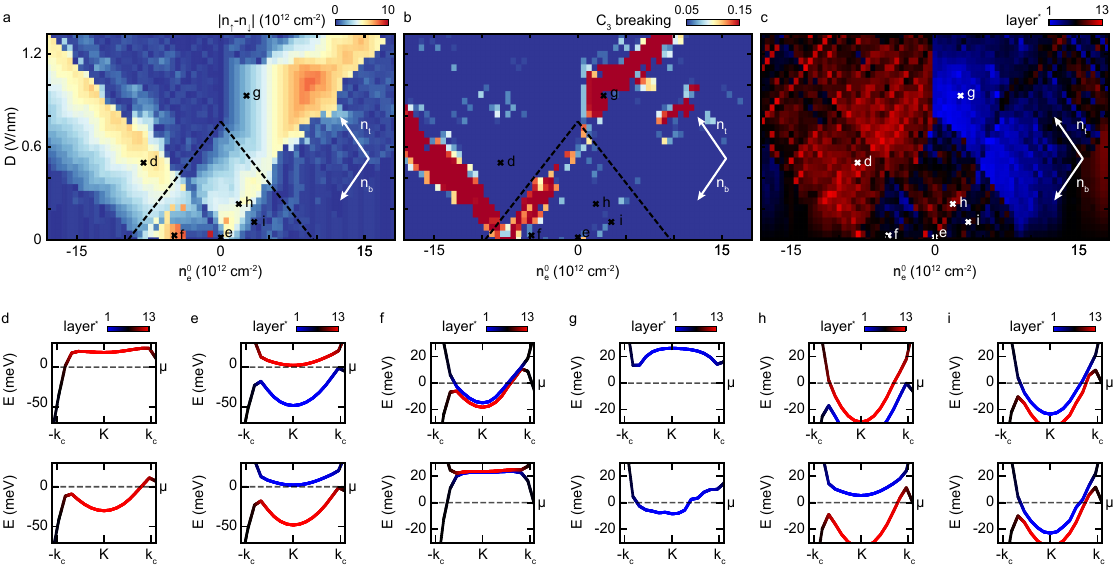}
\caption{
\textbf{Extended results of a Hartree-Fock calculation for $N=13$ layers assuming double (spin) degeneracy.}
\textbf{(a)} Spin-imbalance density.
\textbf{(b)} Momentum-averaged $C_{3}$ symmetry breaking of the mean-field density matrix.
\textbf{(c)} Average layer number of the states at the Fermi level.
\textbf{(d-i)} Band structures at selected points in the phase diagrams shown in panels \textbf{a-c}. The two panels correspond to the two spin species and
the colorcode encodes the layer number.
}
\label{fig:SI_fock}
\end{figure*}

\clearpage

\begin{figure*}[t]
\centering
\includegraphics[width=\textwidth]{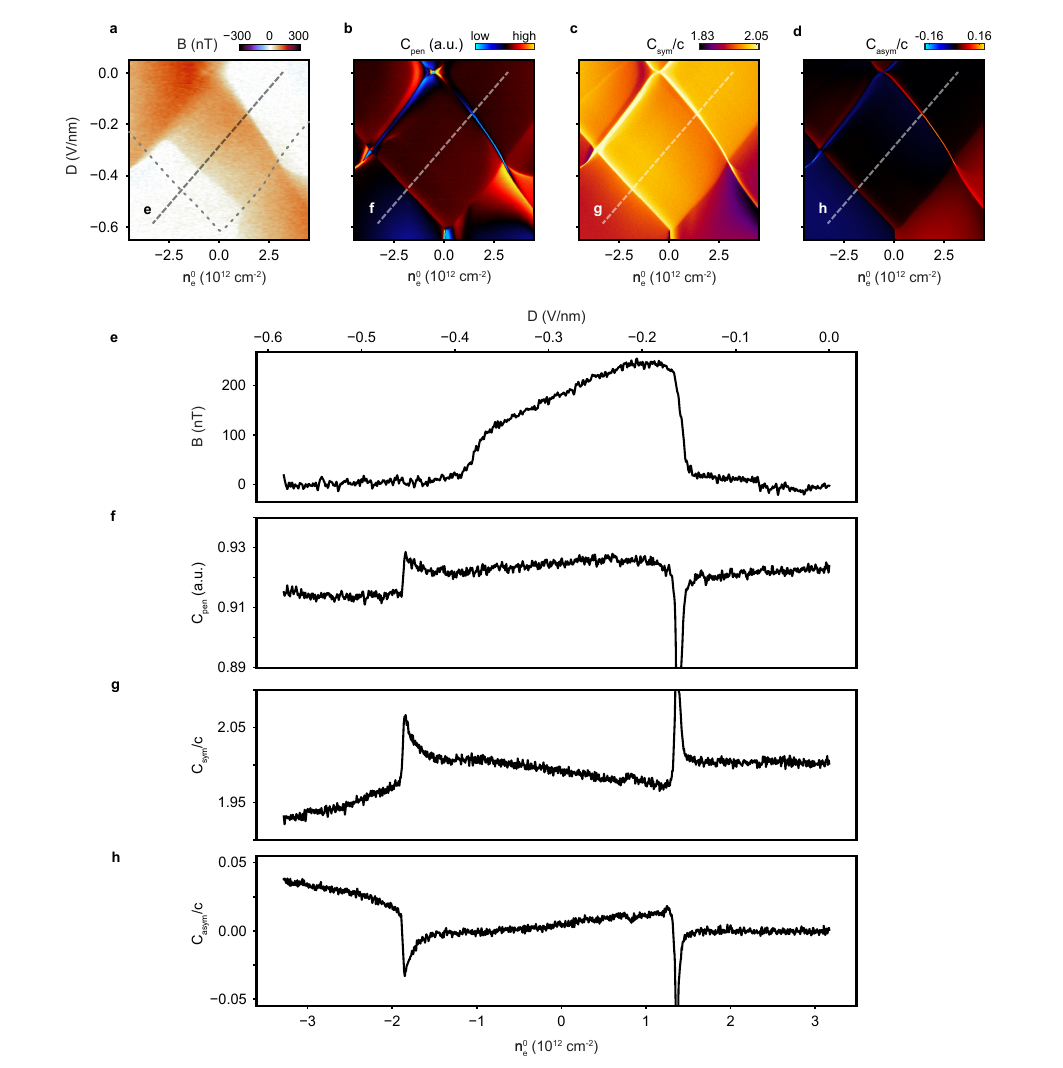}
\caption{
\textbf{Onset of spin polarization.} 
\textbf{(a-d)} Detailed phase diagrams near onset of spin polarization showing 
\textbf{a} local magnetometry, \textbf{b} $C_{pen}$, \textbf{c} $C_{sym}$ and \textbf{d} $C_{asym}$.
\textbf{(e-h)} Line-cuts of the data corresponding to panels \textbf{a-d}, taken along the diagonal (constant $V_{b}$) trajectory from $(n_e^0,D)=(-3.28 \times 10^{12}\,\text{cm}^{-2}, -0.58\,\text{V/nm})$ to $(n_e^0,D)=(3.17 \times 10^{12}\,\text{cm}^{-2}, 0\,\text{V/nm})$. The onset of spin magnetism occurs at a finite doping of the conduction band, inside the dual surface region.
}
\label{fig:SI_SPOnset}
\end{figure*}

\clearpage


\begin{figure*}[t]
\centering
\includegraphics[width=\textwidth]{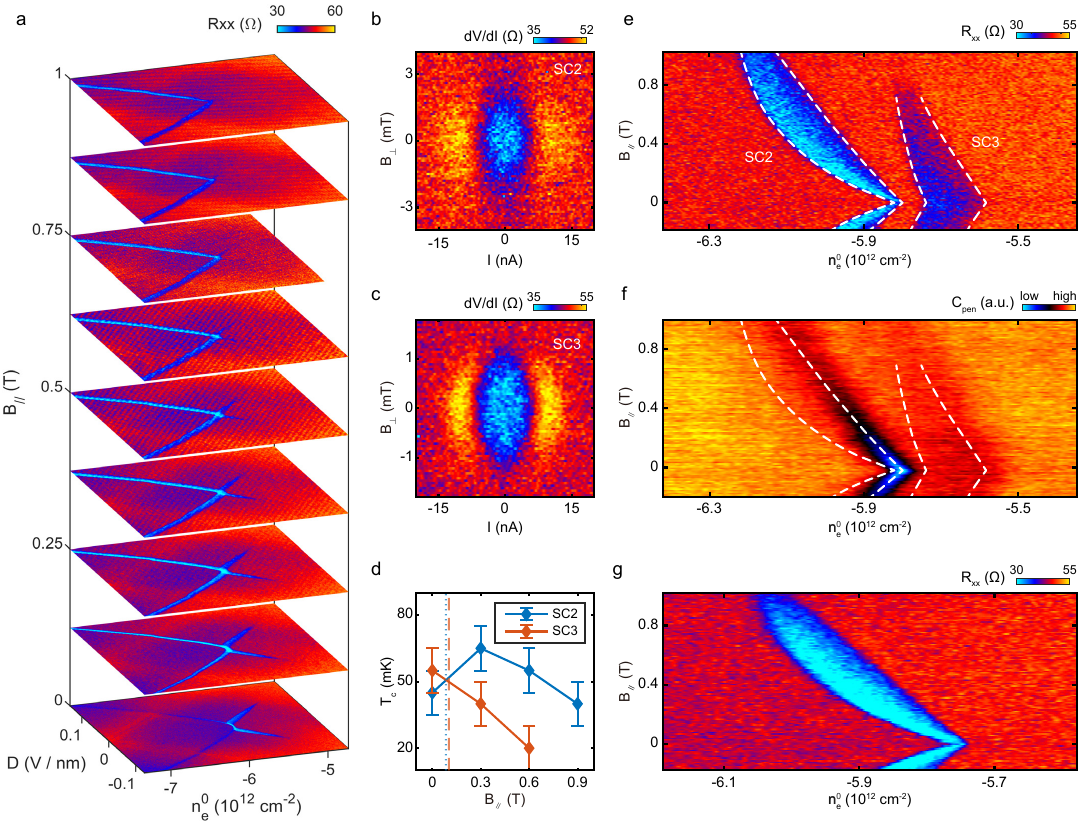}
\caption{
\textbf{Characterization of superconducting states SC2 and SC3.}
\textbf{(a)} $B_\parallel$ dependence of $R_{xx}$ near SC2 and SC3.
\textbf{(b,c)} $dV/dI$ as a function of $B_\perp$ and $I_{DC}$, taken at $D = 0.02\,\text{V/nm}$, $n_e^0 = -5.82$ and $-5.73 \times 10^{12}\,\text{cm}^{-2}$ for SC2 and SC3, respectively.
\textbf{(d)} Critical temperature $T_c$ of SC2 and SC3 as a function of $B_{\parallel}$, where $T_c$ is defined by the onset of the resistivity drop. The dashed lines indicate the Pauli limiting field $B_p=1.86*T_c$, calculated using the zero-field $T_c$.
\textbf{(e)} $R_{xx}$ and \textbf{(f)} $C_{pen}$ as a function of $n_e^0$ and $B_{\parallel}$, measured at $D=0.02 \,\text{V/nm}$. Dashed lines outline the regions of SC2 and SC3.
\textbf{(g)}
$R_{xx}$ as a function of $n_e^0$ and $B_{\parallel}$, measured at $D=0$. 
}
\label{fig:SI_SC23}
\end{figure*}

\clearpage


\begin{figure*}[t]
\centering
\includegraphics[width=\textwidth]{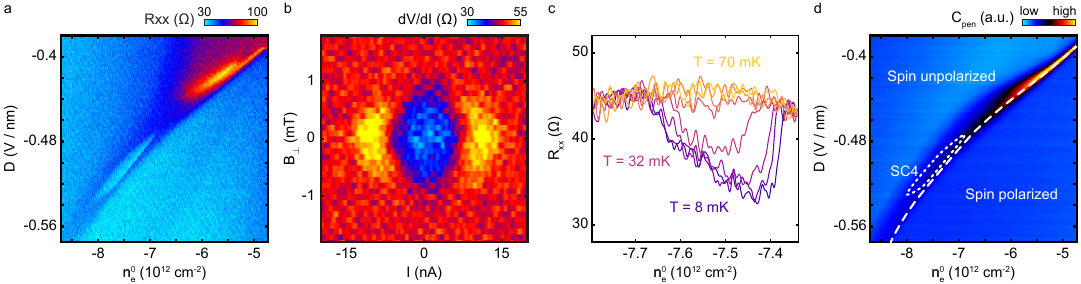}
\caption{
\textbf{Characterization of superconducting state SC4.}
\textbf{(a)} $R_{xx}$ as a function of $n_e$ and $D$. 
\textbf{(b)} $dV/dI$ as a function of $B_\perp$ and $I_{DC}$, demonstrating a critical current of $I_c \approx 7 \,\text{nA}$ and a critical field of $B_c \approx 1 \,\text{mT}$.
\textbf{(c)} Temperature dependence of $R_{xx}$ taken at $D= -0.50~V/nm$.
\textbf{(d)} $C_{pen}$ as a function of $n_e$ and $D$. Dashed lines mark the region of SC4 identified from panel a, and the phase boundary between the full-metal and spin-polarized states.}
\label{fig:SI_SC4}
\end{figure*}

\clearpage


\begin{figure*}[t]
\centering
\includegraphics[width=\textwidth]{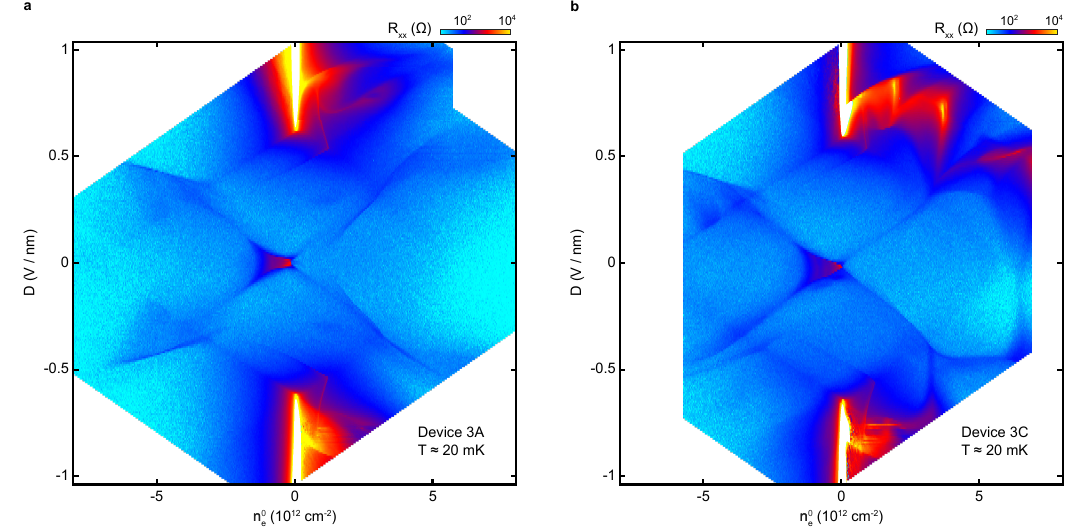}
\caption{
\textbf{Transport phase diagram of 14-layer rhombohedral graphene, measured at $T \approx 20 \,\text{mK}$.}
\textbf{(a)} $n_e-D$ phase diagram of $R_{xx}$ for the pristine 14-layer rhombohedral graphene (Device 3A).
\textbf{(b)} $R_{xx}$ for the 14-layer rhombohedral graphene (Device 3C), crystal axis aligned with the hBN.
}
\label{fig:SI_R14G}
\end{figure*}

\clearpage


\begin{figure*}[t]
\centering
\includegraphics[width=\textwidth]{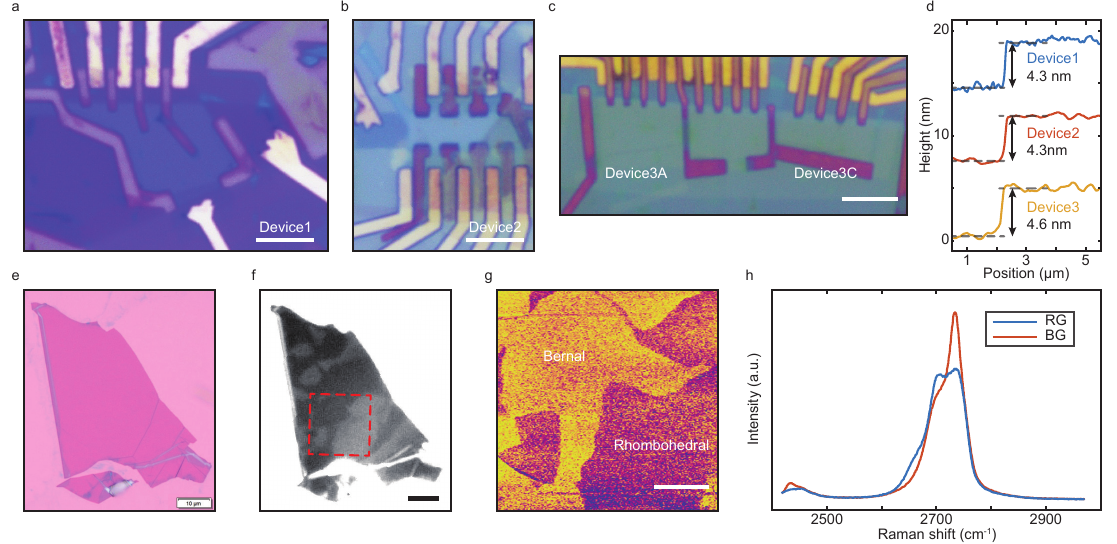}
\caption{
\textbf{Optical micrographs of devices and characterization of rhombohedral stacking order.}
\textbf{(a)-(c)}  Optical microscope images of 13-layer Devices 1 and 2 and 14-layer Device 3, respectively. Scale bar: \SI{5}{\micro\meter}.
\textbf{(d)} Thickness of the rhombohedral graphene flakes contact-mode AFM. The result of $d \approx4.3 \,\text{nm}$ is used to assign the layer number.
\textbf{(e)} Optical and \textbf{(f)} infrared images of the graphene flake used in Device 1. Rhombohedral and Bernal domains show different contrast in infrared image. Scale bar: \SI{10}{\micro\meter}.
\textbf{(g)} Scanning Microwave Impedance Microscopy (SMIM) map taken within the region outlined by the red dashed box in (f). The contrast in microwave impedance distinguishes between Bernal and rhombohedral stacking, revealing the fine structural features of the domains. Scale bar: \SI{5}{\micro\meter}.
\textbf{(h)} Raman spectra of the rhombohedral and Bernal stacking orders.
}
\label{fig:SI_device}
\end{figure*}

\end{document}